\documentclass[a4paper,11pt]{article}
\usepackage{pos}
\usepackage{makecell,bm}
\usepackage{lipsum,afterpage,refcount}

\title{Hadronic Re-Acceleration at the Crab Pulsar Wind Termination Shock as a Source of PeV Gamma-Rays}
 \ShortTitle{Hadronic Re-Acceleration in the Crab Nebula}

\author*[a,b]{Samuel T. Spencer}
\author[a]{Alison M.W. Mitchell}
\author[c]{Brian Reville}

\affiliation[a]{Friedrich-Alexander-Universit{\"a}t Erlangen-N{\"u}rnberg, Erlangen Centre for Astroparticle Physics, Nikolaus-Fiebiger-Str. 2, D 91058 Erlangen, Germany}

\affiliation[b]{Department of Physics, University of Oxford, Clarendon Laboratory, Parks Road, Oxford, OX1 3PU, United Kingdom}

\affiliation[c]{max Planck Institut f{\"u}r Kernphysik, Saupfercheckweg 1, 69117 Heidelberg, Germany}

\emailAdd{samuel.spencer@fau.de}

\abstract{Recent results from LHAASO and Tibet AS$\gamma$ suggest that the Crab Nebula’s gamma-ray spectrum extends to the PeV energy range, however the production mechanisms of this highest energy emission remain unclear. It has been postulated that a secondary component of hadronic emission could explain the highest energy gamma-ray flux points, however the origin and acceleration mechanism for this hadronic population has yet to be explained. We postulate one scenario in which hadrons diffuse over time into the Crab pulsar wind nebula from the surrounding supernova ejecta, and are subsequently re-accelerated by the pulsar wind termination shock. We present results of direct particle transport simulations (including radial evolution) to determine if this scenario is viable over the lifetime of the Crab system.}

\ConferenceLogo{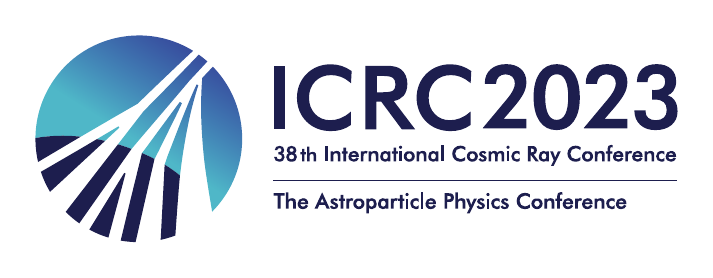}

\FullConference{%
38th International Cosmic Ray Conference (ICRC2023)\\
  26 July - 3 August, 2023\\
  Nagoya, Japan}

\begin{document}
\maketitle

\section{Introduction}
The Crab Nebula is the most widely studied object in Very-High-Energy (VHE) gamma-ray astrophysics \cite{dirsonhorns,zhang,porth}. It is generally accepted that the primary emission mechanism for gamma-rays above a TeV is Inverse Compton (IC) scattering of photons from a variety of background fields (primarily the Cosmic Microwave background at photon energies above $\mathrm{100\,TeV}$) by electrons accelerated at the termination shock of the relativistic wind of the pulsar \cite{LHAASOCRAB}. The recent detection of PeV gamma-ray emission from the Crab Nebula by LHAASO and limits above $\mathrm{100\,TeV}$ from Tibet AS$\gamma$ \cite{LHAASOCRAB,tibet} have re-opened a debate about whether there could be a secondary hadronic component producing the highest energy photons \cite{LHAASOCRAB,LiuWang}, which could potentially be observable due to the Klein-Nishina effect suppressing IC emission at the highest energies \cite{dirsonhorns}. This scenario has not been excluded to date, and despite previous studies (e.g. \cite{LiuWang}) modelling the observations from LHAASO, they have not explained the origin of the PeV particles required. The hadronic fraction of matter in the pulsar wind itself is unknown, but its charge density is constrained to not exceed the Goldreich-Julian density \cite{GJ69}. Alternatively, hadronic material may enter the Pulsar Wind Nebula (PWN) from the outside, though transport through the nebula requires the particles to be already energetic \cite{LucekBell, BellLucek}. The re-acceleration of cosmic rays that enter the PWN, seeded by the supernova remnant (SNR) shocks, has also been explored \cite{ohira}. In that work the re-acceleration follows as a consequence of the crushing of the PWN by the reverse shock of the SNR, but the Crab system is too young ($\sim 1000$ years) for this scenario to apply. 



Global Magneto-Hydrodynamic (MHD) simulations of the Crab Nebula reveal a complex magnetic field topology in the enclosed PWN \cite{porth}. We consider the possibility that protons and other nuclei accelerated at the forward shock of the SNR populate a reservoir of energetic particles that fill the shell surrounding the PWN. In a simple advection versus diffusion picture, particles of sufficiently high energy have a finite probability of traversing this PWN to the pulsar Wind Termination Shock (WTS). At which, these protons could then be re-accelerated, subsequently returning to the outer shell where target material resides. The latter is revealed by the finger-like structures observed in the IR that are believed to be caused by Rayleigh-Taylor instabilities at the PWN boundary. This suggests material is encroaching into the PWN from the surrounding SNR \cite{hester}. Our goal is to determine if the above proposed scenario can account for the highest energy LHAASO flux points.

\section{Method}

We perform particle transport simulations to solve the spherically symmetric transport equation in the PWN, using a Stochastic Differential Equation approach \cite{achterberg1992}. The evolution of the PWN and SNR radius follow the solution of McKee and Truelove \cite{mckeetruelove}; the WTS radius is fixed as $5\%$ of the PWN radius at each timestep. Particles are injected on the evolving interface, at $r=R_{\mathrm{PWN}}$, as the region between the PWN and the rest of the SNR interior is assumed to be a uniform reservoir of cosmic rays. A fixed number of pseudo-particles (1000) are injected at each timestep, distributed as a power-law between energies $T=1\,\mathrm{TeV}$ and $T_2=100\,\mathrm{TeV}$. The normalisation of the pseudo-particles is selected assuming a fraction $\eta$ of the energy of the supernova $E_{\mathrm{SN}}$ has been converted to protons above $\mathrm{1\,GeV}$. Each pseudo-particle is selected from a uniform power-law distribution, \(dN/dE \propto  E^{-S}\), with weighting factor $\alpha$ at injection
\begin{equation}
\alpha=4\pi R_{\mathrm{PWN}}^2 (S-1)\left(\frac{\eta E_{\mathrm{SN}}}{\mathrm{1\,GeV}}\right)\left(\frac{ v_{\mathrm{PWN}} \Delta t}{\mathrm{V_{Res}}(t)}\right)\,,
\end{equation}
where $v_{PWN}$ is the velocity of the outer radius of the PWN as a function of time given by $dR_{\mathrm{PWN}}/dt$, $S$ is the spectral index of the injected protons and $\mathrm{V_{Res}}$ the volume of the particle reservoir between the PWN and SNR forward shock (also given by \cite{mckeetruelove}). We assume a diffusion coefficient with Bohm scaling (in cgs units) 
\begin{equation}
\kappa = \frac{1}{3}\beta r_g c = \beta\left(\frac{Tc}{3ZeB_{\mathrm{max}}}\right)\,,
\label{eq:kappa}
\end{equation}
where $r_g$ is the particle gyroradius, $T$ is the particle energy, $Z$ is the atomic number (1 in all cases) and $B_{\mathrm{max}}$ is the magnetic field strength set to $112\,\mathrm{\mu G}$ (which with single-zone models can reproduce the synchroton and IC emission from X-ray wavelengths to PeV at the current epoch \cite{LHAASOCRAB}). $\beta$ is a correction factor  to account for the mean free path relative to the Bohm limit; for the purposes of these proceedings it is taken as 1. The particle transport equation we solve is given by
\begin{equation}
\frac{\partial F}{\partial t}= -\nabla_Z \cdot [(\dot{\mathbf{Z}} F(\mathbf{Z},t) - \nabla_Z \cdot (\bm{\kappa} F(\mathbf{Z},t))]\,,
\end{equation}
where $\mathbf{Z}$, $F(\mathbf{Z},t)$, $\mathbf{U}=d\mathbf{Z}/dt$,$\nabla_Z$ and $\bm{\kappa}$ are the position vector, particle distribution, velocity, gradient and diffusion tensor in phase space and $\dot{\mathbf{Z}}=\mathbf{U}+\nabla_Z\cdot \bm{\kappa}$ is the effective velocity including a drift term due to diffusivity gradients \cite{schure}. Neglecting synchrotron losses for protons, and assuming spherical symmetry of the shocked wind flow $V_W$ with constant density
\begin{equation}
V_W(r)=\frac{c}{3}\left(\frac{R_{\mathrm{WTS}}}{r}\right)^2,
\end{equation}
results in there being no change in the particle energy at each timestep unless the particle hits the central WTS. In each timestep, the pseudo-particle's radius $r$ changes as 
\begin{equation}
\Delta r = \left(V_W + \frac{2\kappa}{r}\right)\Delta t + \xi_R \sqrt{2\kappa \Delta t}\,,
\label{eq:deltar}
\end{equation}
where $\xi_R$ is a random number following a standard normal distribution centred at 0. We neglect relativistic corrections. For the divergence free velocity field assumed above the particle's energy changes only if it hits the WTS. If the particle hits the shock, its energy is doubled, as expected at an ultra-relativistic shock \cite{achterberg2001} and its updated position is reflected downstream such that $r_{\mathrm{new}}=R_{\mathrm{WTS}}+|R_{\mathrm{WTS}}-r_{\mathrm{old}}|$. Note that for Bohm diffusion, the effective radial velocity for any particle on the shock surface is (for Bohm limit)
\begin{equation}
|\dot{Z}_r| = \frac{c}{3} + \frac{2r_g}{3R_{\mathrm{WTS}}} < c \mbox{~for~} r_g<R_{\mathrm{WTS}}\,.
\end{equation}
Note that $r_g = R_{\mathrm{WTS}}$ corresponds to the Hillas limit for relativistic shocks \cite{hillaslimit}. Thus, provided the time step is chosen such that $\sqrt{2\kappa \Delta t} < c\Delta t$, the maximum energy cannot exceed the Hillas limit, since radial outward directed advection must exceed the diffusive step. For our adopted values, the Hillas limit is $T_{\mathrm{Hillas}}\approx 10^{16}$\,eV.

\begin{table}
\caption{Parameter values used for this particle transport simulation.}
\label{table:params}      
\centering          
\begin{tabular}{c c c c}     
\hline
Parameter & Description & Value & Reference\\ \hline
$t_0$ & Simulation start time & 9 years & -\\
$t_{\mathrm{end}}$ & Simulation end time & 969 years & \cite{crabage}\\
$B_{\mathrm{max}}$ & Maximum magnetic field strength in PWN & $\mathrm{112\,\mu G}$ & \cite{LHAASOCRAB}\\
$\Delta t$ & Timestep & 0.01 years & \cite{LiuWang} (Constraint)\\
$M_{\mathrm{ej}}$& Mass ejected in supernova & $3 M_{\odot}$ & \cite{ohira}\\
$E_{\mathrm{SN}}$ & Supernova energy & $\mathrm{10^{51}\,erg}$ & \cite{ohira}\\
$\eta$ & Fraction of $E_{\mathrm{SN}}$ in protons & 0.00005 & \cite{zhang} (Constraint)\\
$E_0$ & Proton normalisation energy & $\mathrm{1\,GeV}$ & -\\
$L_{\mathrm{sd}}$ & Spin-down luminosity of Crab pulsar & $\mathrm{3\times 10^{38}\,erg\,s^{-1}}$& -\\
$n_{\mathrm{ISM}}$ & Proton density in ISM & $\mathrm{0.10\,cm^{-3}}$ & \cite{ohira} \\
$n_{\mathrm{target}}$ & Proton density in target material & $\mathrm{5\,cm^{-3}}$ & - \\
$D$ & Distance to Crab Nebula & $\mathrm{1.999\,kpc}$ & \cite{kaplan} \\
$\beta$ & Diffusion coefficient relative to Bohm & 1 & -\\
$n_{\mathrm{inject}}$ & Pseudo-particles injected per-timestep & 1000 & - \\
$T_1$ & Minimum pseudo-particle injection energy & $\mathrm{1\,TeV}$ & - \\
$T_2$ & Maximum pseudo-particle injection energy & $\mathrm{100\,TeV}$ & - \\
$S$ & Pseudo-particle injection spectral index & 1.2 & - \\

\hline
\end{tabular}
\end{table}
Only particles shocked at least once are tracked when they escape the PWN. The gamma-ray emission from the target region is then modelled using the GAMERA package \cite{GAMERA} with the cross-section parameterisations of Kafexhiu et al. \cite{Kafexhiu}; the source region is treated as a static target at approximately 1000 years after the simulation. The proton escape spectrum is re-normalised by scaling the flux points relative to a total energy content as a requirement of GAMERA's input.

\section{Results}

\begin{figure}
    \centering
    \includegraphics[width = \hsize]{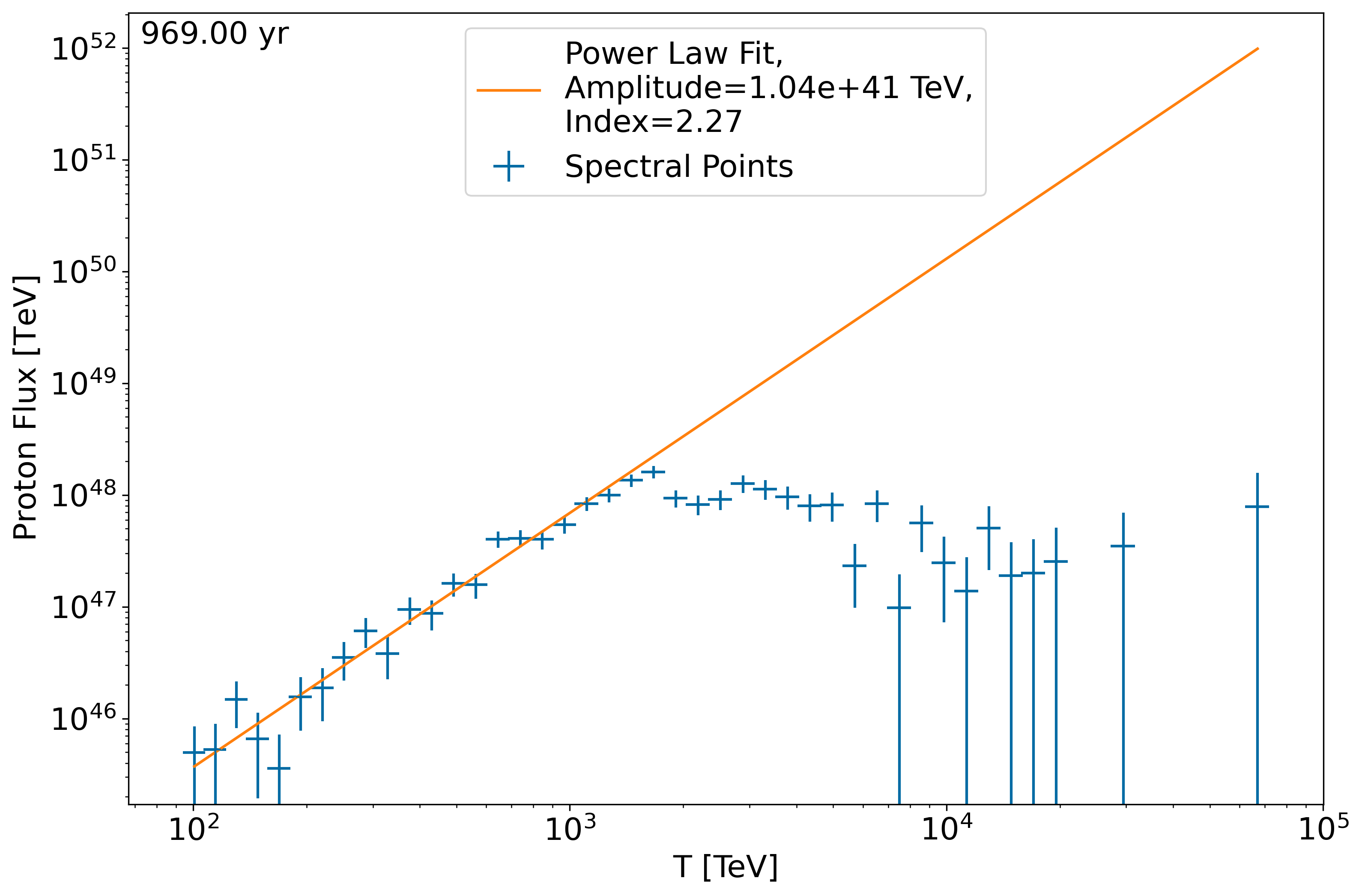}
    \caption{Distribution of re-accelerated protons escaping the PWN region, this is re-normalised prior to injection into GAMERA. A linear fit to the spectrum below $\mathrm{1\,PeV}$ is also shown.}
    \label{fig:finalescape}
\end{figure}

\begin{figure}
    \centering
    \includegraphics[width = \hsize]{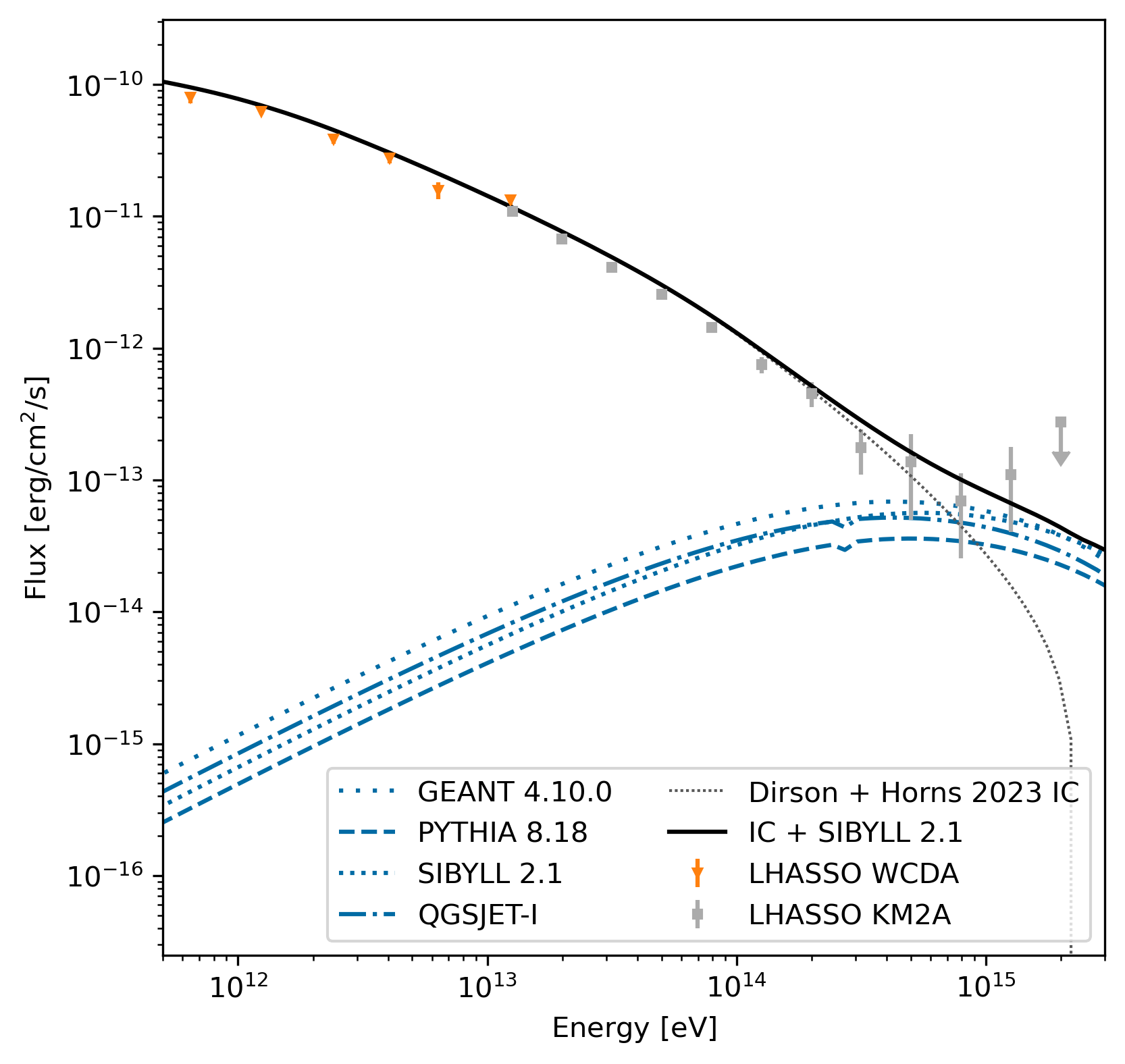}
    \caption{Resulting gamma-ray spectrum of the Crab nebula including the secondary population of hadrons, the multi-instrument Inverse Compton scattering fit from Dirson and Horns \cite{dirsonhorns} is also shown for comparison.}
    \label{fig:spectrum}
\end{figure}

The flux of escaped shocked protons at the end of the simulation, the resulting gamma-ray spectra from the target region, and the particle distribution inside the simulation at the final timestep are shown in Figures \ref{fig:finalescape} and \ref{fig:spectrum}. 
A variety of different hadronic interaction models are explored when using the distribution shown in Figure \ref{fig:finalescape} as an input to GAMERA; SYBIL 2.1 is chosen for the gamma-ray emission fit shown in Figure \ref{fig:spectrum}, which is combined with a multi-band IC model taken from Dirson and Horns \cite{dirsonhorns}. This is as it produces the most optimistic behaviour in terms of having a relatively lower hadronic flux at low energies which then rises at higher energy. For this scenario to be viable, the diffusion coefficient 
has to scale roughly linearly with energy in order for particles to be shocked and subsequently escape. The majority of particles are also shocked relatively early in the PWN's life; it is likely the spectral break at $\mathrm{\sim1\,PeV}$ we observe is a direct result of the energy dependent transport.
The particles that travel the complete distance from $R_{\mathrm{PWN}}$ to $R_{\mathrm{WTS}}$ and back again are very much in the minority, only $\sim700$ of the $\sim10^8$ total injected pseudo-particles do so, but this is a necessary as to not over-estimate the hadronic flux at lower energies. The density assumed for the target region and the fraction of supernova energy in the hadronic population are degenerate in their effect in scaling the normalisation of the hadronic spectrum, to obtain a reasonable quality fit we assume $5\times10^{-5}$ of the supernova energy goes into this population of shocked protons. However, the maximum energy of the protons accelerated at the proton shock we assume ($100\,\mathrm{TeV}$) is arguably optimistic, and the effect of changing this free parameter in the model requires further investigation. There also exists the possibility of there being a small population of accelerated protons in the pulsar wind itself, which we do not consider in this work.


\section{Conclusion}
Our results show that the posited scenario of protons diffusing inwards from a source region between $R_{\mathrm{PWN}}$ and $R_{\mathrm{SNR,fs}}$ to be re-accelerated at $R_{\mathrm{WTS}}$ is feasible, and could explain the PeV emission observed from the Crab PWN by LHAASO. Our plans for future research are to continue investigating this scenario for older sources to see if other hadronic PeVatrons could be detectable with next-generation experiments, and to investigate whether using this model for hadronic re-accleration in PWN could account for other gamma-ray sources detected at energies greater than 100\,TeV.

\section{Acknowledgements}
This work is supported by the Deutsche Forschungsgemeinschaft (DFG, German Research Foundation) – Project Number 452934793.

%
%
%

\end{document}